\documentclass[fleqn,10pt]{wlscirep}
\usepackage[utf8]{inputenc}
\usepackage[T1]{fontenc}

\usepackage{lineno,hyperref}
\usepackage[ruled,vlined]{algorithm2e}
\usepackage{graphicx}
\usepackage{caption}
\usepackage{subcaption}
\usepackage{makecell}
\usepackage{pdfpages}

\modulolinenumbers[5]

\title{Categorising Fine-to-Coarse Grained Misinformation: An Empirical Study of the COVID-19 Infodemic}

\author[1]{Ye Jiang}
\author[1,*]{Xingyi Song}
\author[1]{Carolina Scarton}
\author[1,2]{Ahmet Aker}
\author[1]{Kalina Bontcheva}
\affil[1]{University of Sheffield, Department of Computer Science, Sheffield, S1 4DP, UK}
\affil[2]{University of Duisburg-Essen, Department of Computer Science and Applied Cognitive Science, Duisburg, 47057, Germany}

\affil[*]{Corresponding author: x.song@sheffield.ac.uk}

\begin{abstract}
The spreading of COVID-19 misinformation over social media already draws the attention of many researchers. According to Google Scholar, about 26000 COVID-19 related misinformation studies have been published to date. Most of these studies focus on 1) detection and/or 2) analysing the characteristics of COVID-19 related misinformation. However, the study of the social behaviours related to misinformation is often neglected. In this paper, we introduce a fine-grained annotated misinformation tweets dataset including social behaviours annotation (e.g. comment or question to the misinformation).  The dataset not only allows social behaviours analysis but is also suitable for both an evidence-based or non-evidence-based misinformation classification task. In addition, we introduce 'leave claim out' validation in our experiments and demonstrate that misinformation classification performance could be significantly different when applying to real-world unseen misinformation.

\end{abstract}

\begin{document}

\flushbottom
\maketitle
\thispagestyle{empty}

%\linenumbers

\section{Introduction}

The ubiquity of social media has increasingly affected information dissemination and consumption during the COVID-19 pandemic \cite{sharma2020covid}. With increasing social distancing and growing reliance on online communication, more and more people tend to use social platforms to seek information during the outbreak of COVID-19 \cite{zhou2021characterizing}. However, the complexity surrounding the pandemic not only comes from the virus itself, but also from the surge of the social and behavioural issues that the disease has brought about. While social media allows people to seek information more effectively, the explosion of misinformation has also caused significant harm to the global community. For instance, the misinformation accompanying the outbreak of COVID-19 has caused 1) the deaths of more than 700 people from drinking denatured alcohol \cite{mehrpour2020toll}; 2) doctors being attacked because of misinformation e.g. a WhatsApp user claimed ``health workers were forcibly taking away Muslims and injecting them with the coronavirus''\cite{disinfo}; 3) several 5G towers being burned down by misinformation claiming they cause COVID-19 \cite{bbc_news}. 

Although international fact-checking outlets have increased 400\% since 2014 in 60 countries \cite{Stencel}, false claims and online misinformation are still pervading social platforms. For instance, websites spreading misinformation had almost four times as many estimated views as equivalent content from reputable organisations on Facebook (\url{https://secure.avaaz.org/campaign/en/facebook\_threat\_health/}). To combat COVID-19 misinformation, the worldwide fact-checkers and media are overloaded. The World Health Organisation (WHO) has termed the situation a global infodemic \cite{who} and launched the `Mythbuster' (\url{https://www.who.int/emergencies/diseases/novel-coronavirus-2019/advice-for-public/myth-busters}) platform to fight the spread of COVID-19 related misinformation. However, such counter measures are limited in their ability to combat misinformation, due to the large-scale nature and fast-paced evolution of online discourse \cite{sharma2020covid}. While fact checking by professionals (e.g., International Fact-Checking Network (IFCN) -- \url{https://www.poynter.org/ifcn/}) is a vital defence in the fight against misinformation, it has limited volume and may not reach the intended audience \cite{micallef2020role}. Automatic methods are therefore a pressing need to reliably detect misinformation on a large scale; however this typically requires a large amount of annotated data to model the semantic feature and discourse in online misinformation.

Existing COVID-19 misinformation studies mainly focus on the identification of online misinformation \cite{cui2020coaid, gupta2021global,zhou2020recovery, hossain2020covidlies}, but the occurrences of other social behaviours related to the misinformation are omitted. Understanding the correlation between misinformation and its related questions or comments is crucial to investigate the prevalence of misinformation; particularly since the number of people using social media to ask questions or leave comments related to health advice, especially during health emergencies, is constantly growing \cite{zhou2021characterizing}. Consequently, this paper aims to address three questions: 1) How many social media posts are questioning or commenting about a misinformation claim and do they correlate? 2) Does the volume of tweets debunking the misinformation claim correlate with the volume of misinformation tweets? 3) What are the different kinds of misinformation spreading on social media? 

Meanwhile, previous COVID-19 misinformation research has investigated the social engagement with fake news on websites and social platforms \cite{cui2020coaid}, the ways that misinformation is countered in tweets \cite{micallef2020role} and the stances detection between tweets and misconceptions \cite{hossain2020covidlies}. However, there is a lack of research that evaluates debunk information (provided by professional fact-checkers) as well as a shortage of effective machine learning classifiers (as these debunk information and provide useful references to related misinformation).

Addressing such requirements, in this paper, we introduce:
\begin{enumerate}
\item An information retrieval experiment to retrieve COVID-19 tweets that are related to the IFCN fact-checked misinformation. The fact-checked misinformation claims are used as the queries to extract tweets with topics that are similar to the misinformation. 
\item A manually annotated fine-grained COVID-19 misinformation Twitter dataset with 8 categories that are suitable for training machine learning models to automatically detect misinformation and social behaviours related to COVID-19 misinformation.
\item A quantitative analysis of the fine-grained categories throughout a 10-month period and particularly investigating different kinds of misinformation tweets.
\item A benchmark experiment evaluating the misinformation classification performance of state-of-the-art NLP models over fine-to-coarse grained scenarios. Specifically, the fine-grained classification enables the identification of misinformation and its related social behaviours (e.g. debunks, questions and comments, etc). The coarse-grained classification labels could be reorganised as (a) \textbf{Evidence based misinformation classification task} and (b) \textbf{Non-evidence based misinformation classification task}. In the first task, we aim to detect the misinformation that has already been debunked (the debunked misinformation that is provided by IFCN as the evidence). The misinformation prediction must be supported with verified misinformation. In the non-evidence based task, we aim to find social media posts that are likely to be misinformation; however these posts may require human verification. 
\end{enumerate}

According to our previous findings \cite{song2021classification} the topics of COVID-19 misinformation change significantly in different stages of the pandemic. Evaluating classification over unseen topics is an important step to assess the model performance in a real-world situation. In this paper, our tweets are retrieved according to IFCN claims (as queries). Therefore, the annotated tweets are organised to its associated claim. In our experiment, we introduce 'leave claim out' cross-validation (CV) to ensure the topics between training and testing data are different. Hence to simulate the real world classification performance over unseen misinformation. To the best of our knowledge, we are the first to introduce 'leave claim out' cross-validation to evaluate the misinformation classification performance over unseen claim/topics. We also compare the 'leave claim' CV with the 'leave claim out' CV. Our experimental results suggest that misinformation classification performance is significantly reduced when applying to real-world unseen misinformation.

\section{Related Work}
Misinformation on social media has been an increasingly pervasive problem in recent years \cite{sharma2019combating}. The dissemination of misinformation related to civil discourse \cite{allcott2017social}, natural disasters \cite{gupta2013faking} and health emergencies \cite{zhou2021characterizing} has been studied in different social contexts. Given the popularity of Twitter in the global community, past research has highlighted the importance of studying Twitter during epidemics. For example, 255 million Twitter users were found active in February 2014 during the start of the Ebola outbreak \cite{Tw1} and this number topped 330 million in 2019 \cite{Tw2}. Therefore, Twitter can be utilised as a rich source for the research community to study the prevalence of online misinformation during the COVID-19 pandemic.

\subsection{COVID-19 Dataset}
With the outbreak of COVID-19, several datasets have been established to assist research communities to fight the pandemic. Singh et al. \cite{singh2020first} investigate the early conversations about the pandemic on Twitter, and analyse five predefined myths as well as links to poor quality tweets between January and March 2020. Dong et al. \cite{dong2020interactive} establish a real-time tracking of COVID-19 to help epidemiological forecasting. Chen et al. \cite{chen2020tracking} collect COVID-19 scholarly articles for literature-based discoveries, and track the information spread on Twitter. To analyse how social behaviours are affected by the outbreak of COVID-19 and the spread of related information on social media on a large scale, Lamsal \cite{781w-ef42-20} collected 310 million English language tweets related to COVID-19 and analysed the sentiment, relations between countries and hashtags. Gruzd and May \cite{SP2/PXF2CU_2020} release a multi-lingual Twitter dataset with around 270 million tweets. Gupta et al. \cite{gupta2021global} retrieve over 132 million tweets from around 20 million unique user IDs, and investigate their latent topics, sentiment and emotions by applying topic models and sentiment analysis.

In terms of datasets that particularly focused on misinformation related to COVID-19, Micallef et al. \cite{micallef2020role} investigate the tendency of the misinformation and counter-misinformation (aka. debunks) tweets based on two predefined topics (i.e. Fake Cures and 5G Conspiracy Theories). These datasets focus on predefined topics and themes, but topics of COVID-19 misinformation are fast-evolving. To tackle this, Cui and Lee \cite{cui2020coaid} (CoAID) combine news articles published by reliable media outlets to identify the misinformation on Twitter. Sharma et al. \cite{sharma2020covid} label tweets as misinformation if the tweet shared any article or content posted from any of the misinformation sources compiled using the fact-checking sources. However, it is hard to measure the reliability of such data since there is no gold-standard annotation. Saakyan et al. \cite{saakyan2021covid} (COVIDFACT) introduced a `Counter-Claim' algorithm that automatically generated false COVID-19 related claims based on the subreddit r/COVID19 discussion, and obtains a moderate agreement of 0.47 for contradictory claims between models and humans. In this paper, we use professional IFCN verified claim for misinformation classification, but we are planning to introduce automatic generated false claims as an additional source to speed up the misinformation debunking. Hossain et al. \cite{hossain2020covidlies} (COVIDLIES) divide misinformation detection into two sub-tasks: 1) relevant tweets retrieval based on COVID-19 misconceptions, and 2) stance detection to identify whether the tweets agree or disagree with the misconceptions. Several automatic methods are evaluated based on a manually annotated dataset, but the data time span is restricted to only a one-month period (i.e. from March to April 2020), meaning the assessment of long term tendencies of misinformation is not possible. Compared with the above dataset, our dataset investigates a longer time span over 10 months, which cover tweets from the first and second wave of outbreaks in the US and UK. We also use debunks which were provided by the professional fact-checkers (which provide evidence for the misinformation tweets).

\subsection{COVID-19 Misinformation Detection}
Singh et al. \cite{singh2020first} define five Coronavirus Common Myths based on some keywords searching on the websites. Then the misinformation is identified based on phrases and words in the tweets and from broad descriptions of the myths (taken from the original searches that described each myth). Sharma et al. \cite{sharma2020covid} compile information from three fact-checking sources (i.e., Media Bias/Fact Check, NewsGuard and Zimdars) that provide journalistic analysis of low-quality news sources known to frequently publish unreliable and false information. Similarly, Zhou et al. \cite{zhou2020recovery} apply Media Bias/Fact Check and NewsGuard to filter out news sites that are reliable/unreliable, and track misinformation based on the URLs and user information in the tweets. 

Several studies apply machine learning methods to model the semantic feature in the misinformation. Micallef et al. \cite{micallef2020role} train three one-vs-all Logistic Regression classifiers to automatically identify misinformation, counter-misinformation and irrelevant tweets respectively. Cui and Lee \cite{cui2020coaid} evaluate the hierarchical attention network (HAN) \cite{yang2016hierarchical} and its variant dEFEND \cite{shu2019defend} on the CoAID datasets. Song et al. \cite{song2021classification} propose a classification-aware neural topic model for a COVID-19 disinformation category classification and topic discovery. Meanwhile, Li et al. \cite{li2021exploring} evaluate the pre-trained language model BERT \cite{devlin2018bert} with ensemble techniques on the AAAI 2021 Shared Task: COVID-19 Fake News Detection in English. Hossain et al. \cite{hossain2020covidlies} combine BERTScore \cite{zhang2019bertscore} with Sentence BERT to identify tweets stance for COVID-19 related misconceptions. However, those misinformation detection methods do not evaluate the effectiveness of using debunk information provided by the professional fact-checkers. In this paper, we investigate how the debunks would potentially effect the misinformation detection performance.

\begin{figure*}[h!]
\centering
\includegraphics[scale=0.7]{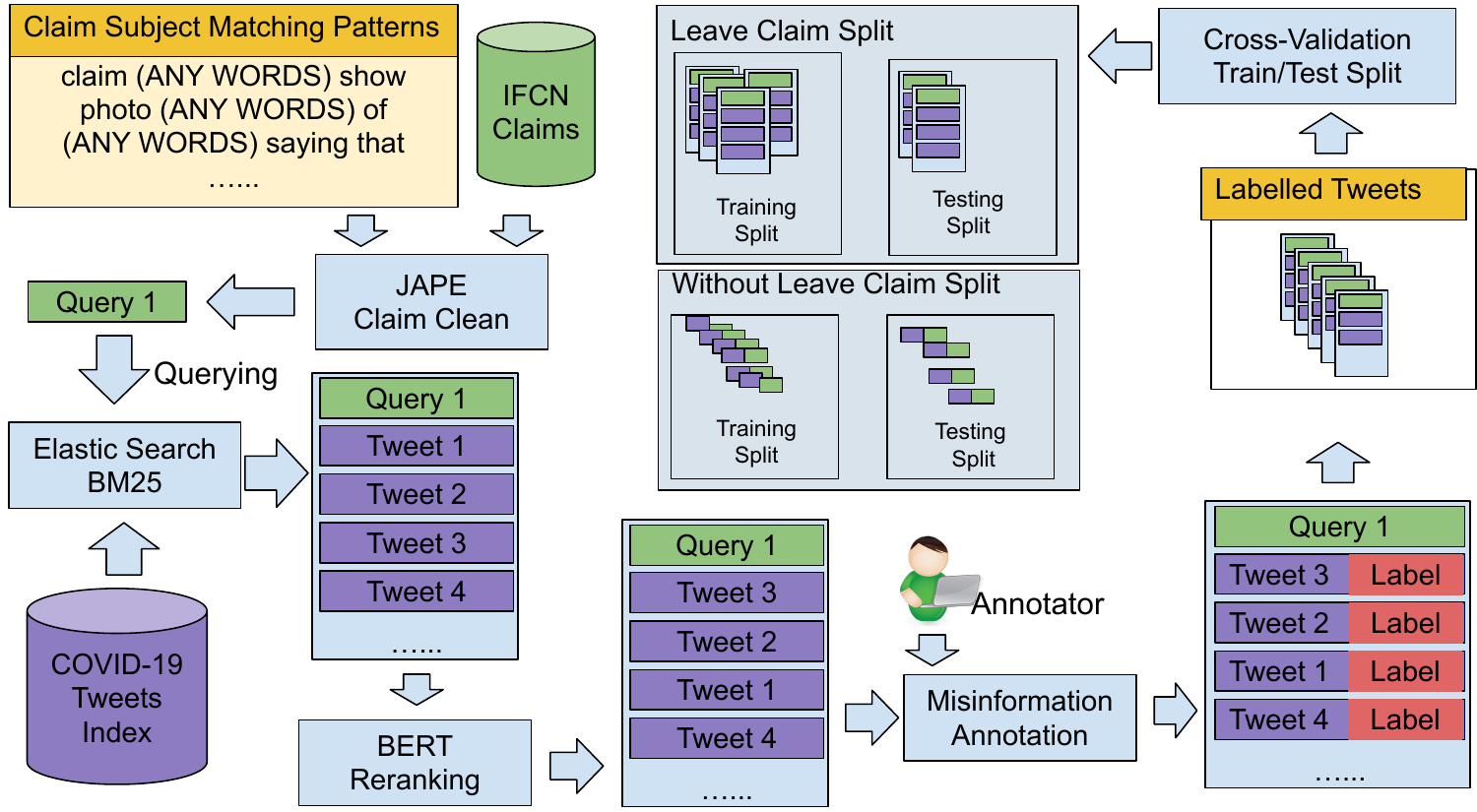}
\caption{Overall pipeline}\label{fig:pipeline}	
\end{figure*}
 
\section{Dataset and Annotation}
The overall pipeline of dataset annotation is shown in Figure \ref{fig:pipeline}. In general, we first collect tweets based on a set of keywords and create a COVID-19 related tweets index into the Elasticsearch (https://www.elastic.co/elasticsearch/). We also create a fact-check dataset, which includes fact-checked misinformation and its meta data from the IFCN websites, and select 90 misinformation claims as the queries to retrieve related tweets from the index. In order to improve the relevance of tweets, similar to previous research \cite{hossain2020covidlies}, we implement a Transformer-based model to re-rank the retrieved tweets based on their semantic similarities. The re-ranked tweets are then annotated based on fine-grained categories, and the agreement rates between annotators are evaluated. Finally, several classification tasks are conducted based on different types of data validation methods. 

\subsection{Tweet Collection}
We first identify a collection of keywords (e.g, covid, covid-19, coronavirus, covid\_19, etc.) related to COVID-19 and collect tweets that contain one of those keywords in the hashtag. We use the Twitter Stream API (\url{https://developer.twitter.com/en/docs/tutorials/consuming-streaming-data}) to collect 182,027,646 English tweets  spanning 10 months from March to December 2020. Then, we create ElasticSearch index for the tweets that are collected.

\subsection{IFCN Dataset}

In order to have a fact-checked list of COVID-19 related misinformation, we also build a IFCN dataset by utilising the work of fact-checkers. First, we extract 10,381 fact-checked misinformation claims (referred to as `claims' in the remaining parts of the paper) from the IFCN Poynter website (\url{https://www.poynter.org/ifcn-covid19-misinformation/}). We select 90 English claims from April 2020, focusing on claims that appeared in the UK and US, since we wanted to maximise the number of tweets in English that could be retrieved. The IFCN claim extraction and process steps follow the same procedures as our previous research  \cite{song2021classification}
A pattern matching language -- JAPE \cite{cunningham2000jape} is applied to remove the subject from the claim in order to obtain a precise expression of the misinformation. e.g. ``\textit{Japanese doctor who won Nobel Prize said coronavirus is artificial and was manufactured in China}'' the subject ``Japanese doctor who won Nobel Prize said'' is removed and the claim shortened to ``\textit{coronavirus is artificial and was manufactured in China}''. The example subject patterns used in this work can be found in Figure~\ref{fig:pipeline} `Claim Subject Matching Patterns' (yellow) box. 

\subsection{Tweets Retrieval and Re-ranking} \label{sec:retrieval}
The selected 90 IFCN claims are used as the queries to retrieve tweets from the Elasticsearch index. The initial retrieval utilises BM25 algorithm \cite{robertson1995okapi} to extract the 1,000 most relevant tweets from the Twitter index. To mitigate the cost of retrieval time, we then implement a tinyBERT \cite{jiao2019tinybert} model, which has been pre-trained based on the MS MACRO dataset \cite{nguyen2016ms} for document ranking, to re-rank the retrieved tweets based on the semantic similarities between queries and tweets. After re-ranking, we select the 20 most relevant tweets for each misinformation, based on the cosine similarity scores. In addition, we restrict the retrieval for tweets posted in a date range of 10 weeks before and 2 weeks after the debunk date. This way, we aim to collect tweets related to a specific misinformation in a certain time, since similar misinformation can appear at different stages (e.g. misinformation about generic topics like 'a nurse in Italy died after taking the COVID-19 vaccine' may appear and re-appear at different times, in different countries, depending on the the vaccine rollout). 

\subsection{Annotation}
We obtained 1,800 tweets after the initial retrieval and re-ranking. Nine volunteer annotators were recruited and we gave them the instructions available in Appendix \ref{app:guidelines} for annotating tweets. The fine-grained categories are listed as following:
\begin{enumerate}
    \item \textbf{Misinformation}: Tweets contain falsehoods, inaccuracies, rumours, decontextualised truths, or misleading leaps of logic, and deliver exactly the SAME information/topic as the claim.
    
    \item \textbf{Related Misinformation}: Tweets contain falsehoods, inaccuracies, rumours, decontextualised truths, or misleading leaps of logic, and deliver a SIMILAR information/topic with the claim but towards, for instance, a different person name, event name, medication name, illness name, etc.

    \item \textbf{Debunk}: Tweets refute exactly the SAME information/topic as the claim, and are generated either by professional fact-checkers e.g.government website, IFCN, etc.,  or general citizen responses with/without use of any checkable evidence e.g. reputable links, hashtags, etc.
    
    \item \textbf{Related Debunk}: Tweets refute a SIMILAR information/topic with the claim but towards, for instance, a different person name, event name, medication name, illness name, etc.,  and are generated either by professional fact-checkers e.g. government website, IFCN, etc.,  or general citizen responses with/without use of any checkable evidence e.g. reputable links, hashtags, etc. 
    
    \item \textbf{Question}: Tweets raise a question based on the exact SAME information/topic as the claim.
    
    \item \textbf{Comments}: Tweets add some comments on the exact SAME information/topic as the claim.
    
    \item \textbf{Relevant Others}: A tweet is not misinformation or a debunk of the claim but is nevertheless about the topic of the given claim.
    
    \item \textbf{Irrelevant}: The information/topic of the Tweets that are IRRELEVANT to the claim.
\end{enumerate}

Before the formal annotation, a pilot annotation was conducted so as to train the annotators. The formal annotation task was then conducted in a 3-weeks period. We created groups with three annotators each and we kept the same annotators in each group throughout the 3-weeks task, so each entry was annotated three times to evaluate the annotation agreements. Each annotator was assigned 200 tweets in each week. 

During annotation, each entry provided to the annotators presented the query, the date when the misinformation was debunked, the fact-checkers’ explanation, the organisation who fact-checked the misinformation, the misinformation veracity (e.g. false, misleading), and the source link to the fact-checkers’ own web page. 
The volunteers assign each tweet with the most relevant of the eight fine-grained categories, and indicate their confidence (on a scale of 0 -- least confident -- to 5 -- most confident) as well as their comments, if any. The tweet ID, the tweet text, the link to the tweet, and the date of when the tweet was posted were also provided.

We calculate the Krippendorff's alpha for each week to assess the data quality, and the final averaged score among the three weeks is 0.67, which demonstrates a substantial agreement between annotators. The final dataset is produced by merging the multiple-annotated tweets on the basis of: 1) majority agreement between the annotators where possible; or 2) confidence score, if there was no majority agreement, the label with the highest confidence score was adopted. From the 1,800 tweets, 78 tweets did not have either majority agreement or a valid confidence score, so we removed those tweets in the final dataset. The statistics of the final annotated dataset are shown in Table \ref{tb:anno_stat} and examples of tweets in each class can be found in Appendix \ref{app:example}. 

\begin{table}[h]
\begin{center}
\begin{tabular}{|c|c|c|c|c|}
\hline
Misinformation &  Related Misinformation & Debunk & Related Debunk & Question\\
\hline
522 & 175 & 194 & 56 & 115\\
\hline
Comment & Irrelevant & Relevant Others & Total & \\
\hline
99 & 199 & 362 & 1722 & \\
\hline
\end{tabular}
\end{center}
\caption{Number of examples per category in the final dataset. \label{tb:anno_stat} }
\end{table}

\subsection{Data Analysis}

\begin{figure*}[h!]
\centering
\includegraphics[scale=0.35]{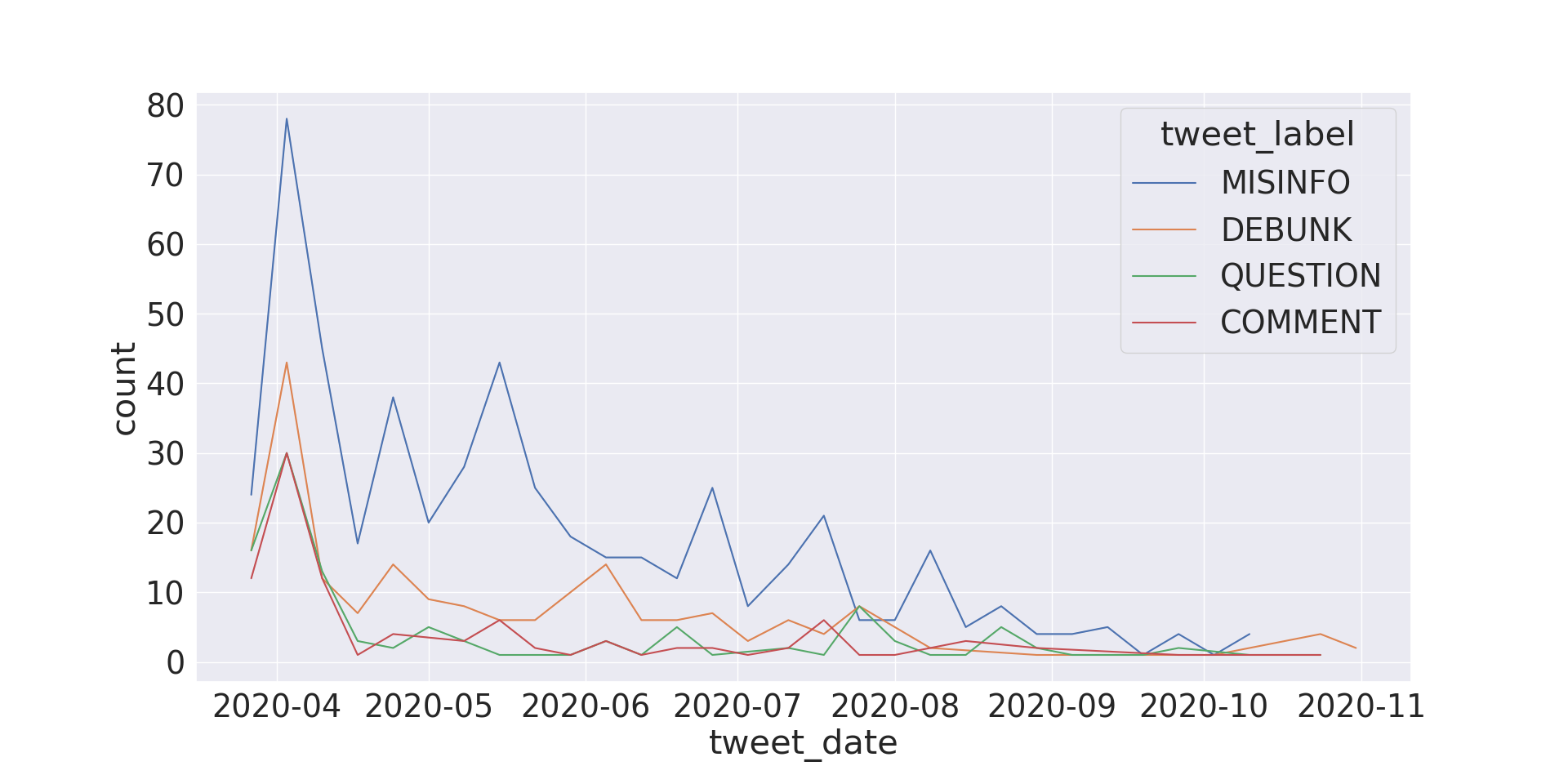}
\caption{Misinformation, debunk, question and comment tweets volume over time (in weeks). }\label{fig:tweet_volume}	
\end{figure*}

One of the aims of this work is to understand the correlation of the spread of misinformation and debunk with other behaviours. As shown in Figure \ref{fig:tweet_volume}, firstly, the volume of misinformation tweets is significantly higher than the other categories, especially at the beginning of April, which coincides with when the first wave of the pandemic started in both United State and United Kingdom. Secondly, there is a significantly higher volume of 'question and comment' tweets indicating that people tend to seek information and leave comments at the beginning of the first wave, but this tendency is decreasing throughout the pandemic. We also observe that there is a notable correlation between misinformation and debunk tweet counts (Pearson correlation $\rho=0.55$, $p<0.001$). This indicates that misinformation tweets and debunk tweets are spread at the same rate, similar to the findings made in \cite{micallef2020role} and \cite{mendoza2010twitter}. The misinformation tweets also have a positive correlation with comment tweets (Pearson correlation $\rho=0.58$, $p<0.001$) and question tweets (Pearson correlation $\rho=0.45$, $p<0.001$), this is similar to the debunk tweets with comment tweets (Pearson correlation $\rho=0.54$, $p<0.001$) and question tweets (Pearson correlation $\rho=0.41$, $p<0.001$).
Overall, we find that the debunk tweets have a similar spreading rate as misinformation tweets. In addition, people tend to leave comments or ask questions when there is a significantly high number of misinformation and debunks tweets.

\subsubsection{URL Sources in Misinformation and Debunk Tweets}

\begin{figure*}[h!]
\centering
\includegraphics[scale=0.35]{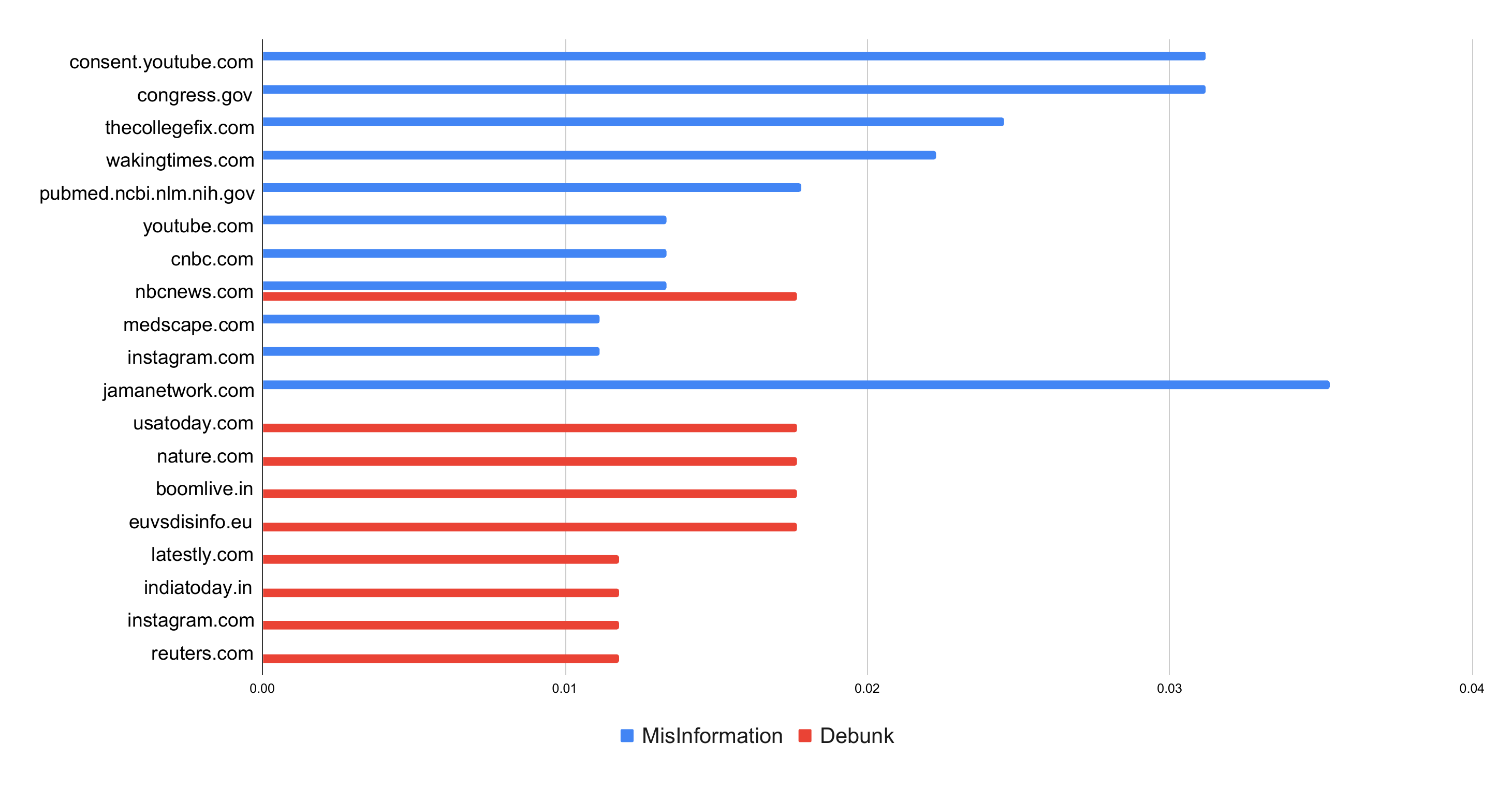}
\caption{Top 10 frequent URLs found in misinformation and debunk tweets. }\label{fig:tweet_url}	
\end{figure*}

The top 10 frequent URL domain names found in misinformation and debunk tweets shown in Figure~\ref{fig:tweet_url}. The numbers in horizontal axe are averaged by number of misinformation/debunk tweets. We note that there is almost no URL overlap between misinformation and debunk tweets (only overlap URL is cnbc.com), and misinformation tweets are very likely to link to video website (e.g. youtube.com). We also note that misinformation tweets have high frequency contain URLs than that in the debunk tweets, and may also contain high-credibility sources (e.g.PubMed).  For instance, a misinformation tweet claims that \textit{`Now officially : 5G Technology and induction of coronavirus in skin cells published online ahead of print, 2020 Jul 16. J Biol Regul Homeost Agents, 2020'} and provides a link to `pubmed.ncbi.nlm.nih.gov'. However, that paper was retracted after a thorough investigation as it showed evidence of substantial manipulation of the peer review. In addition, several tweets quote information from \textit{`clinicaltrials.gov'} and claim that \textit{`Hydroxychloroquine and Zinc With Either Azithromycin or Doxycycline for Treatment of COVID-19 in Outpatient Setting'}. However, large-scale clinical trials demonstrate no beneficial effect of hydroxychloroquine in terms of viral shedding, disease severity, or mortality among COVID-19 patients. 

\subsubsection{Hashtags in Misinformation and Debunk Tweets}
Similarly to URLs, we note that a 'hashtag' is a strong indicator to misinformation as well as debunk tweets. We found that some misinformation hashtags have negative emotion towards a person or an organisation (e.g., EvilGates, FireFauci, etc.) and some are generally denying the pandemic (e.g., FakePandemic, coronascam, etc.). On the other hand, hashtags in debunk tweets are less emotional (e.g., FactMatter, SeekReliableSource, etc.), and some directly indicate the professional fact-checkers or high-credibility source (e.g., AltNewsFactCheck, pubmed, PIBFactCheck, etc.). Wordclouds of misinformation and debunk tweets can be found in Appendix \ref{app:wordclouds}.

\section{Experiments} \label{sec:exp}
In this section, we conduct a benchmark experiment for our annotated Twitter misinformation classification dataset. This experiment includes three tasks that represent three different misinformation classification scenarios. The task detail and the experiment settings are discussed in Section~\ref{sec: tasks}. Then, we introduce the baseline models and model configurations in Section \ref{sec:model}. Finally, the experimental results are discussed in Section \ref{sec:results}.

\subsection{Misinformation Classification Tasks}

The classification experiment is divided into three tasks. Besides the fine-grained classification task, which takes account of all labels based on the evidences, we also introduce two coarse-grained classification tasks according to the different hierarchy methods of the fine-grained classes. The descriptions of each task are listed in the following paragraphs, and the corresponding labels for coarse-grained non-evidence based and evidence-based classification tasks are illustrated in Table~\ref{tb:labelcate}.

\label{sec: tasks}
\begin{table}[h]
\begin{center}
\begin{tabular}{|l|l|l|}
\hline
\multicolumn{3}{|c|}{Coarse-grained Evidence Based Classification} \\
\hline
\bf{Misinformation} & \bf{Debunk} & \bf{Other} \\
\hline
Misinformation & Debunk   & Comment \\
& & Relevant Other \\
& & Irrelevant \\
& & Related Misinformation \\
& & Question \\
& & Related Debunk \\
\hline
\multicolumn{3}{|c|}{Coarse-grained Non-Evidence Based Classification} \\
\hline \bf{Misinformation} & \bf{Debunk} & \bf{Other}  \\ \hline
Misinformation & Debunk & Question\\
Related Misinformation & Related Debunk  & Comment \\
& & Relevant Other\\
& & Irrelevant \\
\hline
\end{tabular}
\end{center}
\caption{Coarse-grained classification label hierarchy. The bold texts are the coarse-grained labels, and its corresponding fine-grained labels are listed in the column beneath \label{tb:labelcate} }
\end{table}

\begin{enumerate}
    \item \textbf{Fine-grained misinformation classification}: Classify the tweet text into one of the eight fine-grained labels introduced in this paper. This task aims to identify the tweets that might be misinformation, debunk or other associated behaviours (e.g.tweets that leave comments about debunks or tweets that question about misinformation, etc). Since the information/topics of `Misinformation' and `Debunk' tweets are the same as the IFCN claim, and IFCN claims are served a evidences in our classification task, the fine-grained misinformation classification task is therefore evidence based.  
    \item \textbf{Coarse-grained evidence based misinformation classification}: Similar to fine-grained classification, this task aims to classify tweets that have already been debunked, but concentrates more on the misinformation and debunk tweets. In this case, tweets labelled with `Misinformation' will be treated as `\textit{Misinformation}' tweets and tweets labelled with `Debunk' will be treated as `\textit{Debunk}' misinformation. All other labels, including `Related Misinformation/Debunk' are categorised as `\textit{Other}'. 
    \item \textbf{Coarse-grained Non-evidence based misinformation classification}: This task aims to classify tweets likely to be misinformation, where there are no debunks available. Therefore, different to the coarse-grained evidence based task, the `Related Misinformation/Debunk' labels are categorised as `\textit{Misinformation}/\textit{debunks}', together with `Misinformation/Debunk' tweets.
\end{enumerate}

For each classification task, we report the results based on 5-fold cross-validation. The evaluation metrics used in this experiments are 1) accuracy, 2) F1 measure for each class, and 3) macro average F1 (i.e. the average of class level F1 Measure) across all classes.  Two different folding methods are used in this experiment:
\begin{itemize}
    \item Folding {\bf without Leave Claim Out}: This is the standard 5-fold cross-validation. The training data is randomly split into five sub-groups. For each sub-group, one sub-group is retained as the validation set, and the remaining sub-groups are used for training.
    \item Folding {\bf with Leave Claim Out}: Similar to the standard 5-fold cross-validation, but the random sub-group splitting is based on claim rather than on all training data. Therefore no claim in the test set will appear in the training stage. This is a more realistic testing method to test model performance on `unseen' misinformation since most of the online misinformation have not been debunked by the professional fact-checkers in the real world. 
\end{itemize}

\subsection{Model and Configuration}
\label{sec:model}
Four state-of-the-art baseline models are used in this experiment to benchmark the classification task performance. BERT\_CLS and CANTM are the evidence independent models used to test the classification performance without providing claim (please note, claims are applied in this work as evidence) information. BERT\_Pair and SBERT are evidence dependent models and have been widely applied in Natural Language Inference tasks. In this experiment, we apply these two models to test the performance with the aid of evidence information. 
\begin{itemize}
    \item \textit{BERT\_CLS}: The BERT \cite{devlin2018bert} version used in this experiment is a 24 transformer layers (BERT-large) COVID-Twitter pre-trained \cite{muller2020covid} BERT. Only the parameters in the last transformer encoding layer is unlocked for fine-tuning, the rest of the BERT weights were frozen for this experiment. BERT\_CLS treat all tasks as a Tweet text classification task. 
    The model input is [CLS] + Tweet\_Text + [SEP], and the final hidden state of [CLS] token will be the representation of Tweet\_Text. The probability of labels is predicted using a Softmax classifier based on the Tweet\_Text [CLS] representation.
    \item \textit{CANTM}: Classification-Aware Neural Topic Model is a stacked asymmetric variational autoencoder that outputs classification and topic predictions. In this experiment, we only consider the classification output of CANTM model. CANTM apply the BERT model as input text encoder, and the BERT model setting is the same as BERT\_CLS. The vocabulary size for CANTM is 3,000 with 50 latent topics.
    \item \textit{Sentence-BERT} (SBERT): We apply SBERT \cite{reimers2019sentence} classification objective function for our classification experiment. SBERT classification objective function aiming to optimise the cross-entropy loss of a softmax classifier ($o=softmax(W(q,t,|q-t|))$). The input feature of the classifier is the weighted concatenation of evidence embedding ($q$), tweet text embedding ($t$) and the element-wise difference $|q-t|$. In this experiment, all embeddings are obtained from [CLS] token of COVID-Twitter pre-trained \cite{muller2020covid} BERT, and apply the same setting as \textit{BERT\_CLS}. The evidence of the tweet text is the claim that is described in Section~\ref{sec:retrieval}.
    \item \textit{BERT\_Pair}: Similar to BERT\_CLS, but BERT\_Pair also takes evidence into consideration. The input of the model is [CLS] + Evidence + [SEP] + Tweet\_Text + [SEP]. BERT\_Pair has been originally applied for the next sentence prediction task and has been fine-tuned for pair-wise text classification such as Natural Language Inference. The probability of labels is predicted using a Softmax classifier based on the pairwise [CLS] representation. We experiment BERT\_Pair model with two different settings: 1) The results labelled with BERT\_Pair\_{MNLI} are trained with the Multi-Genre Natural Language Inference (MNLI) corpus \cite{N18-1101}. The MNLI labels contradiction, entailment and neutral corresponding to the debunk, misinformation, and other in our misinformation classification task.  2) The results labelled with BERT\_Pair are trained with our labelled misinformation data (5-fold cross-validation)
\end{itemize}

\subsection{Coarse-Grained Classification Results}
\label{sec:results}

Table \ref{tb:norefres} shows the results of coarse-grained misinformation classification tasks. In the without `leave claim out' cross validation all models achieved more than 0.75 accuracy in both evidence- and non-evidence-based classification tasks. The best performed models are SBERT and BERT\_Pair. Both models are evidence dependent and able to reach around 0.8 classification accuracy in both coarse-grained tasks. 

\begin{table}[h]
\resizebox{\textwidth}{!}{\begin{tabular}{|l|r|r|r|r|r|r|r|r|r|r|}
\hline
\multicolumn{11}{|c|}{{\bf Without Leave Claim Out Cross Validation}} \\ \hline
& \multicolumn{5}{|c|}{Non-Evidence-Based Classification Task} & \multicolumn{5}{|c|}{Evidence-Based Classification Task} \\ \hline
 & Acc. & \makecell{Avg. \\ F1} & \makecell{Debunk \\F1} & \makecell{MisInfo \\ F1} & \makecell{Other \\ F1} & Acc & \makecell{Avg. \\ F1} & \makecell{Debunk \\F1} & \makecell{MisInfo \\ F1} & \makecell{Other \\ F1}  \\ \hline
BERT\_CLS       & 0.789 & 0.771 & 0.709 & 0.803 & 0.799 & 0.759 & 0.715 & 0.608 & 0.729 & 0.808 \\
CANTM           & 0.792 & 0.762 & 0.664 & 0.816 & 0.806 & 0.779 & 0.722 & 0.597 & 0.739 & 0.830 \\
SBERT           & 0.808 & 0.789 & 0.724 & 0.815 & 0.828 & 0.804 & 0.753 & 0.643 & 0.765 & 0.851 \\
BERT\_Pair      & 0.797 & 0.787 & 0.749 & 0.807 & 0.804 & 0.808 & 0.757 & 0.665 & 0.760 & 0.846 \\
\hline
  \multicolumn{11}{|c|}{{\bf With Leave Claim Out Cross Validation}} \\ \hline
  
BERT\_CLS       & 0.648 & 0.609 & 0.487 & 0.672 & 0.668 & 0.632 & 0.533 & 0.405 & 0.490 & 0.705 \\
CANTM           & 0.640 & 0.584 & 0.448 & 0.647 & 0.657 & 0.622 & 0.477 & 0.252 & 0.453 & 0.724 \\
SBERT           & 0.662 & 0.613 & 0.476 & 0.681 & 0.681 & 0.632 & 0.550 & 0.409 & 0.526 & 0.715 \\
BERT\_Pair      & 0.634 & 0.595 & 0.470 & 0.656 & 0.657 & 0.643 & 0.567 & 0.468 & 0.508 & 0.724 \\
\hline
BERT\_Pair\_{MNLI} & 0.455 & 0.396 & 0.384 & 0.227 & 0.578 & 0.514 & 0.395 & 0.312 & 0.219 & 0.655 \\
\hline
\end{tabular}}
\caption{COVID-19 coarse-grained misinformation classification results. \label{tb:norefres} }
\end{table}

Compared between two coarse-grained tasks, all baseline models have lower average F1 scores in the evidence-based classification task than non-evidence-based classification. This may be because: 1) \textit{Evidence-based classification is a more challenging task}. In the non-evidence-based classification, the misinformation or debunks can be determined according to previously learned topics/information that was included in the training data. However, evidence-based classification is a pairwise classification task, misinformation/debunks can only be determined according to the given evidence. Hence, a tweet text cannot be classified as misinformation/debunk if it does not match the given evidence even the tweet text is misinformation/debunk (with other evidence). 2) \textit{Data is more imbalanced in evidence-based classification task}. According to the label hierarchy (Table~\ref{tb:labelcate}), related misinformation and debunks are categorised as `Other' class in the evidence-based classification. This reduces the number of training samples in the misinformation/debunks classes, and increases the samples in the other class. According to the results, although the average F1 scores are lower, the `Other' class F1 scores are better than the non-evidence-based classification task.

In the `leave claim out' cross-validation, all models decreased at least 15\% in average F1 measure compared to `without leave claim out' cross-validation. This is expected, since in the `leave claim out' cross-validation, the topics between training and testing set are different, and models cannot make a prediction based on its learned misinformation topics. According to the results, models are over-fitted to the misinformation topics from the training set. This also indicates that keeping the training data up-to-date is important to maintain the model's real-world misinformation classification performance. 

According to the class-level F1 score, the performance of misinformation classification is better than debunk classification. This may happen because of the class imbalance problem. The number of debunk and related debunk samples is much smaller (about $1/3$) than misinformation and related misinformation samples. This problem is also reflected in the number of debunking posts being much smaller than the misinformation posts on social media. A faster misinformation debunk using an automated NLP algorithm will help prevent misinformation.

In the last row of Table~\ref{tb:norefres} the classification performance of the Multi-Genre Natural Language Inference trained BERT\_Pair$_{MNLI}$ model is shown (the average F1 score of MNLI mismatched development set is 0.73). The BERT\_Pair$_{MNLI}$ have almost identical F1 score (0.39) in both tasks. Hence, the traditional natural language inference trained model may not be suitable for misinformation classification.

\subsection{Fine-Grained Classification Results}\label{sec:resultsFine}

Table \ref{tb:finerefres} shows the results of the fine-grained misinformation classification task. The fine-grained misinformation classification task is evidence-based. This task further split the \textit{other} class from the coarse-grained evidence-based classification task into six more granular classes (Related Debunk, Related Misinformation, Comment, Question, Relevant Other and Irrelevant) according to the given evidence. In the `without leave claim out' cross-validation, all models drop around 0.2 average F1 scores compared to the coarse-grained evidence-based classification task. The main performance decrease occurred in the fine-grained `Other' classes. The debunk and misinformation class-level F1 measure remains similar in performance (but slightly worse) as the coarse-grained evidence-based classification task. This is because the number of misinformation and debunk training samples are the same as coarse-grained evidence-based classification. The main challenge of the fine-grained classification is to predict samples from `Other' classes further into six fine-grained classes.

\begin{table}[h]
    \centering
    \resizebox{\textwidth}{!}{\begin{tabular}{|l|r|r|r|r|r|r|r|r|}
\hline
& \multicolumn{4}{|c|}{Without Leave Claim} & \multicolumn{4}{|c|}{Leave Claim} \\
\hline               & BERT\_CLS & CANTM & SBERT & BERT\_Pair & BERT\_CLS & CANTM & SBERT & BERT\_Pair \\ \hline
Accuracy             & 0.584 & 0.621 & 0.639 & 0.615 & 0.310 & 0.349 & 0.353 & 0.370 \\
F1                   & 0.515 & 0.524 & 0.555 & 0.524 & 0.271 & 0.277 & 0.259 & 0.276 \\
Debunk F1            & 0.622 & 0.638 & 0.630 & 0.602 & 0.333 & 0.312 & 0.361 & 0.382 \\
MisInfo F1           & 0.671 & 0.736 & 0.757 & 0.742 & 0.373 & 0.476 & 0.535 & 0.495 \\
R-Debunk F1          & 0.293 & 0.264 & 0.409 & 0.258 & 0.025 & 0.0   & 0.071 & 0.038 \\
R-MisInfo F1         & 0.416 & 0.439 & 0.478 & 0.434 & 0.135 & 0.085 & 0.069 & 0.131 \\
COMM F1              & 0.239 & 0.224 & 0.159 & 0.209 & 0.110 & 0.221 & 0.143 & 0.149 \\
QUES F1              & 0.715 & 0.695 & 0.719 & 0.697 & 0.613 & 0.623 & 0.451 & 0.578 \\
REL F1               & 0.595 & 0.624 & 0.646 & 0.635 & 0.335 & 0.343 & 0.309 & 0.320 \\
IRREL F1             & 0.573 & 0.572 & 0.643 & 0.613 & 0.248 & 0.158 & 0.131 & 0.116 \\
\hline
\end{tabular}}
\caption{COVID-19 misinformation fine-grained query based classification. The corresponding class label are R-Debunk:Related Debunk, R-MisInfo:Related Misinformation, COMM:comment, QUES:question, REL:Relevant Other, IRREL:irrelevant \label{tb:finerefres} }
\end{table}

\begin{figure}[htb!]
\centering
\begin{subfigure}[b]{0.55\textwidth}
\includegraphics[width=\textwidth]{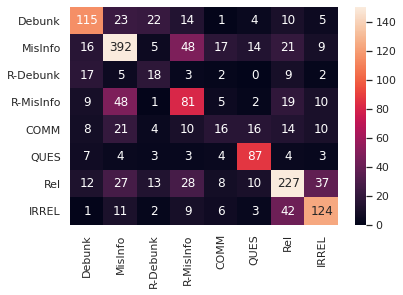}
\caption{} \label{fig:confmatrixnoleave}
\end{subfigure}
\begin{subfigure}[b]{0.8\textwidth}
\includegraphics[width=\textwidth]{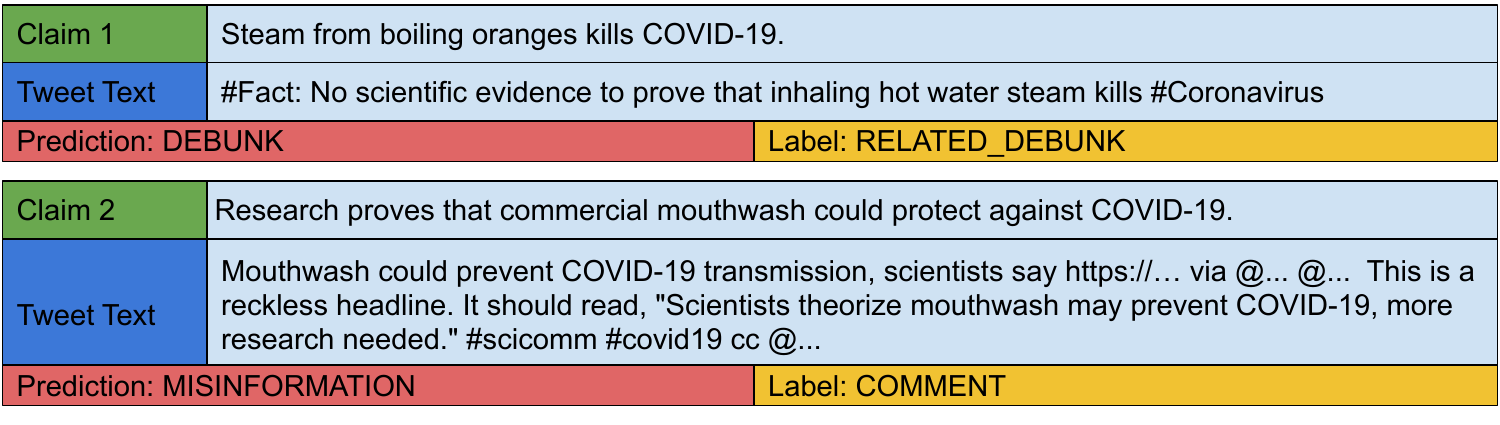}
\caption{} \label{fig:errorexp}
\end{subfigure}
\caption{(a) BERT\_Pair confusion matrix in the fine-grained classification 'without leave claim out' validation. Numbers in each row are the number of samples labelled in the corresponding class, and numbers in each column are the number of samples which have been predicted in the corresponding class. (b) Example of misclassified cases.}	
\end{figure}

Figure~\ref{fig:confmatrixnoleave} shows the confusion matrix of BERT\_Pair results in the fine-grained classification `without leave claim out' validation. According to the figure, most `Related Debunk/Misinformation' samples are misclassified as `Debunk/Misinformation'. This may happen because all training samples are semantically similar to the IFCN claim (the training samples are the top 20 tweets with the highest BERT embedding cosine similarity to claim), and the model is unable to catch the difference between them. An example of this error type is presented in Figure \ref{fig:errorexp}, Claim 1. 

The misinformation claim states that steam from "boiling oranges" kills COVID-19. However, the tweet text being classified is debunking steam from "boiling water" kills COVID-19. The debunk is not directly addressing the query misinformation, therefore, the label should be "RELATED DEBUNK".

Another major classification error occurs in the `Comment' class. The class level F1 scores for the `Comment' class are less than 0.25 with all baseline models. According to the confusion matrix, the `Comment' labelled samples are very likely to be classified as misinformation. The comment class contains tweets that make a comment about the misinformation. Therefore, the misinformation is included in the comment tweet, which might be the main cause of this error. In Figure~\ref{fig:errorexp}, Claim 2 is an example of comment text. The tweet text quote a misinformation claim `Mouthwash could prevent COVID-19 transmission' and make comment that `more research needed' for this claim.

In 'leave claim out' cross-validation, all model average F1 score less than 0.3. Therefore, none of the baseline models are reliable for unseen fine-grained misinformation classification. This may be because all models are over-fitted with training data due to the limited number of samples in most classes. We also note that, only the `Misinformation' class-level F1 score remains similar to the coarse-grained query-based task, and the `Misinformation' class have the most number of samples in the dataset.

\section{Conclusion}
This paper introduced a fine-grained COVID-19 misinformation dataset, which contains 1,722 tweets with eight categories that are manually annotated. In our dataset, each tweet is triple annotated and the averaged Krippendorff's alpha is 0.67 which indicates a substantial agreement. To answer the research question above, we first found that misinformation tweets have similar spread rate to debunk tweets. Secondly, our dataset also enables the investigation of the occurrences of other social behaviours (e.g. questions or comments related to a misinformation) in tweets. We found both question and comment tweets have positive correlation with misinformation and debunk tweets. Thirdly, we also found that misinformation tweets can contain a URL from high-credibility sources. In addition, the hashtags in misinformation tweets are found to be more emotional, and debunk hashtags are more related to the professional fact-checkers. 
Our experiments in Section~\ref{sec:exp} conduct three misinformation classification benchmark experiments: 1) Non-evidence based classification 2) Evidence based classification and 3) Fine-grained classification. The results demonstrate that the all baseline models well performed in standard `without leave claim out' validation across all classification tasks. However, the classification performance dropped significantly with `leave claim out' setting. Therefore, we need to regularly update training instances to ensure the classification performance over time. In the future, we need to develop a classification method to adapt to the fast topic changing nature of misinformation.

\section{Acknowledgements}
This work is funded by the EU H2020 WeVerify (grant agreement: 825297) and SoBigData++ (grant agreement: 871042) projects.

\bibliography{mybibfile}

\newpage

\renewcommand{\figurename}{}
\renewcommand{\thefigure}{S\arabic{figure} Fig}
\setcounter{figure}{0}

\renewcommand{\tablename}{}
\renewcommand{\thetable}{S\arabic{table} Table}
\setcounter{table}{0}

%\setcounter{page}{1}
%\pagenumbering{alph}
\appendix

\section{Annotation guidelines}
\includepdf[pages={1}]{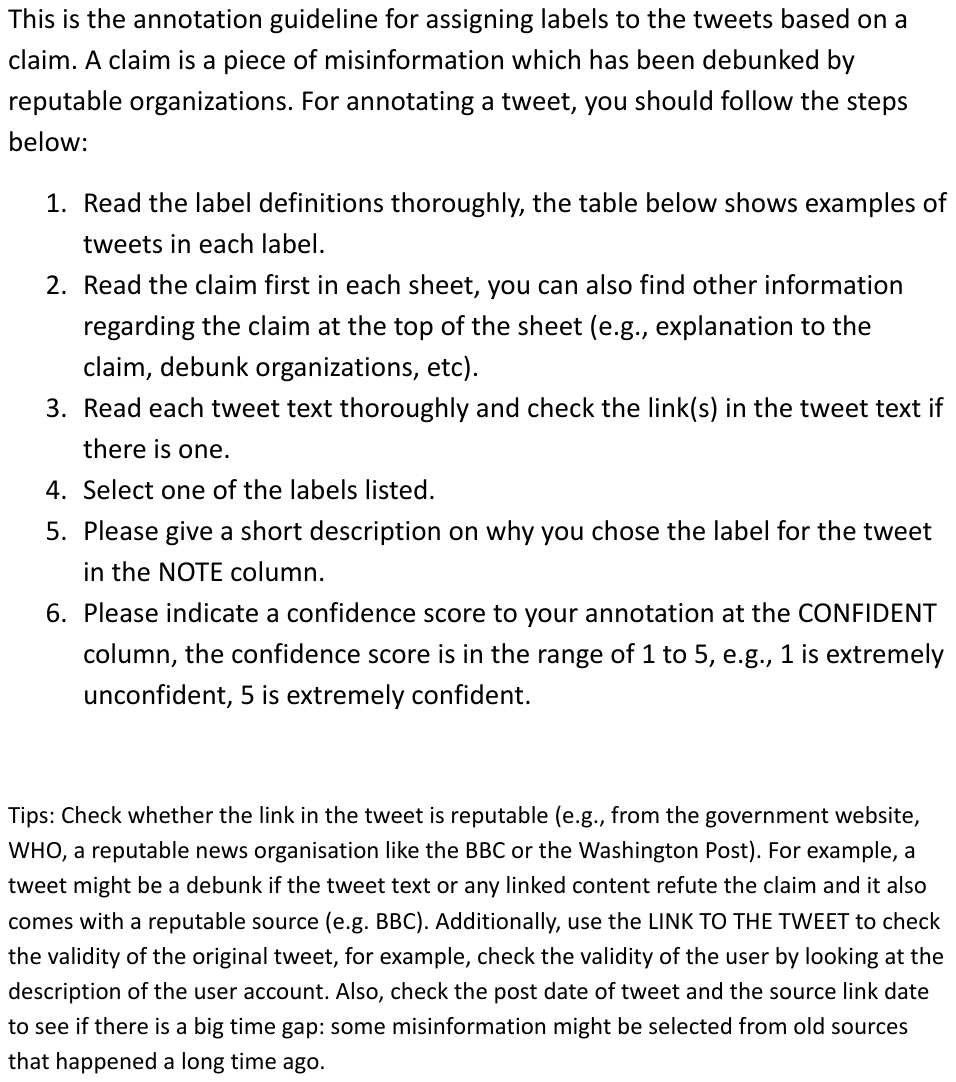}
\includepdf[pages={2}]{figures/Annotation_guidelines.pdf}
\label{app:guidelines}

\section{A example of fine-grained COVID-19 misinformation dataset.}

\label{app:example}
\begin{table}[h!]
\begin{center}
\begin{tabular}{p{\textwidth}}
\hline   
\textbf{Misinformation}: 5G is weakening the immune system, which causes  the coronavirus. \\
\textbf{Related misinformation}: 5g towers cause weakening immune systems.   5g is harmful to human cells. \\
% \textbf{Related misinformation 2}: 5g weakens the immune system by design. Leaving you vulnerable to infection. Symptoms may be similar, but 5g doesn't give you covid19.\\
\textbf{Question}:They say 5G weakens our immune system. Is 5G killing people and animals???\\
\textbf{Comment}: I see a lot of people calling others dumb who believe 5G could contribute to weakening the immune system which creates greater risk to contract Covid-19 I suggest you watch this. \\
\textbf{Debunk}: There’s Zero evidence 5G can weaken immune system or causes Coronavirus COVID19 spread in Iran with Zero 5G coverage  Do not spread panic and fake news! \\
\textbf{Related Debunk}: There’s no evidence 5G is going to harm our health, so let’s stop worrying about it. \\
\textbf{Relevant Other}: Saw the JoeRogan podcast  The question about 5G is, does it weaken the immune system?  There are other questions about this virus too of course That is the 5G one though. \\
\textbf{Irrelevant}: Shelter In Place Is Weakening The Immune Systems Of Everyone Who Complies. \\
\hline
\end{tabular}
\end{center}
\end{table}

\section{Dataset wordclouds from misinformation and debunk tweets.}
\label{app:wordclouds}
\begin{figure*}[h!]
\centering
\includegraphics[scale=0.8]{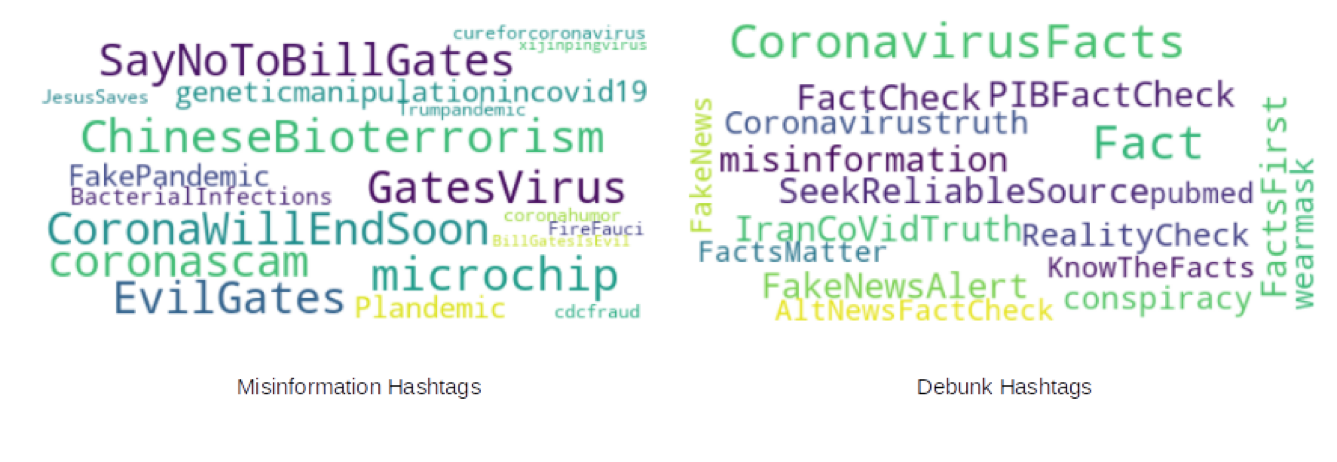}
\end{figure*}

\end{document}